\documentclass[10pt, conference, letterpaper]{IEEEtran}

\usepackage{paralist}
\usepackage[b]{esvect}

\usepackage{caption}
\usepackage{subcaption}
\usepackage{siunitx}
\usepackage{graphicx}
\usepackage{amsmath}
\begin{document}

%

\title{Fast Collision Simulation for Cyclic Wireless Protocols \vspace*{-0.5cm}}

%
%
%
%
%

%

\author{\IEEEauthorblockN{Philipp Kindt, Sangyoung Park, Samarjit Chakraborty}
\IEEEauthorblockA{Chair of Real-Time Computer Systems (RCS), Department of Electrical and Computer Engineering, Technical University of Munich\\
kindt/park/chakraborty (at-sign) rcs.ei.tum.de\vspace*{-0.35cm}}
}

\maketitle
\begin{abstract}
 With most modern smartphones supporting  wireless protocols such as Bluetooth Low Energy (BLE) or ANT+, the number of networks are growing rapidly. Therefore, collisions among multiple networks need to be considered for choosing the appropriate protocol parameters. With growing numbers of networks, simulations for estimating the collision rate become computationally very complex and lengthy. The large simulation times therefore constitute a major limitation in the analysis of complex cases. In this paper, we present a novel simulation technique which can speed up collision simulations by one order of magnitude in realistic situations. Whenever the transmission of a packet is simulated, the cyclic nature of protocols like BLE is exploited to predict the next packet that has a chance of colliding. All transmissions in between can be skipped without affecting the simulation results. Based on the transmission intervals of the networks, one can compute a certain shrinkage per cycle $\gamma$ of the time offset between their packets. Using $\gamma$ and the current offset between the starting times of any two packets, our proposed simulation model can accurately predict the next pair of packets that needs to be simulated. Whereas our proposed technique aims at the BLE protocol, the theory is generic and can be used for many other cyclic protocols such as ANT/ANT+ as well.
\end{abstract}



\section{Introduction}
Mobile ad-hoc networks (MANETS) have become pervasive in our everyday life. In particular, since all modern smartphones have a built-in Bluetooth Low Energy (BLE) radio,
the usage of BLE networks is expected to grow significantly, with 1.2 billion devices predicted to be sold in a single year in 2020 \cite{bleReport:14}. Like in other protocols such as ANT/ANT+ \cite{AntSpec:14}, packets in BLE are exchanged in a purely cyclic fashion with an interval $T_c$. This paper aims towards simulating collisions among multiple networks that communicate using such cyclic protocols.

Whereas BLE systems work flawlessly if only one network is present, packet collisions will become an issue at crowded places with multiple BLE networks in the future. For example, at public squares, at events like concerts and cinemas, in lecture halls, etc., a large number of BLE networks will be within range of reception and therefore interfere with each other. The collision probability is strongly dependent on the protocol configuration (viz, connection interval, packet length and transmit power). In order to minimize collisions in such situations, an estimation of the collision probability is needed. However, in most cases, no analytical model is available for protocols like BLE and ANT+. Existing models can only handle special situations, as we will describe in Section \ref{sec:related_work}.

\begin{figure}[t]
	\centering
	\includegraphics[width=\linewidth]{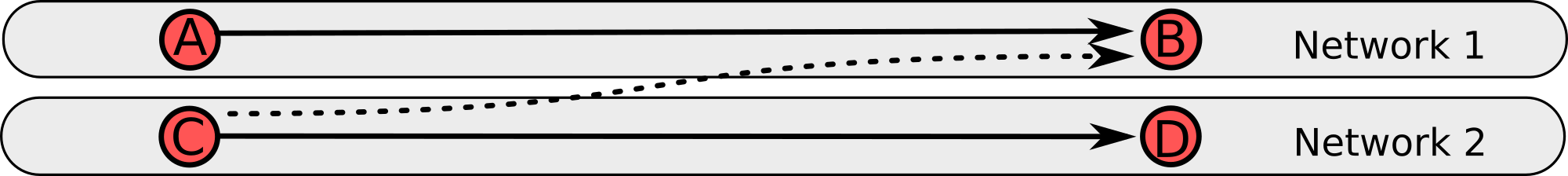}
	\caption{Interfering networks.}
	\label{fig:nodes} 
\end{figure}

Further, even in situations which can be modeled with analytical methods, all existing models do not account for different transmit power levels and the spatial correlation among the nodes. 
Such a situation is exemplified in Figure~\ref{fig:nodes}. Two pairs of nodes (viz. network 1 and network 2) exchange packets between each other, of which some are transmitted on the same channel at the same time.
A collision occurs only if the interfering signal from an adjacent network (network 2) arrives to a receiving node (B) with a sufficient strong power compared to the signal from the sender (A) of network.


To estimate the collision rate of scenarios with more than two interfering nodes or with non-trivial spatial correlations, discrete event simulations are inevitable since analytical models are not available for cyclic protocols. Whereas non-cyclic protocols can be modeled by assuming a certain distribution for the time-offsets between the packets sent, in purely cyclic protocols, these offsets are predetermined by the transmission time of the first packet of each network. Such simulations usually rely on Monte Carlo methods, in which certain parameters (e.g., the initial points in time at which the nodes transmit a packet for the first time) are chosen randomly and the simulation is repeated for many times with different parameters. 
The complexity of the problems under simulation is limited by the available computation time. Even for getting precise estimates on the collision probability among a few BLE networks, billions of discrete events need to be executed.  
For example, a Monte Carlo simulation of the collision rates between two nodes (37 channels, random connection intervals between $7.5~ms$ and $T_{c,m}$ which is increased from $7.5~ms$ to $625~ms$ in steps of $1.25~ms$, 1000 simulation runs per $T_{c,m}$) required the execution of $3.4$ billion events in our experiments. The complexity grows further for increasing numbers of networks, exceeding the computational capabilities that todays computers can handle in a reasonable amount of wall-clock time. Simulations of realistic situations can easily take days or even months.
Hence, it is important to perform simulation very efficiently to evaluate the collision rate.

In this paper, we for the first time present a novel simulation speedup technique, \textit{adaptive event skipping}, which can reduce the mean computation time needed for such a simulation by one order of magnitude in realistic scenarios, and by more than two orders of magnitudes in some special cases. The technique selectively executes simulation events that are potentially related to collisions and skips the rest of the events that are not. The theoretical foundations of this technique are $\gamma$-processes, which allow modeling the shrinkage of the temporal distance between two consecutive, cyclic events. 

The main insight we exploit is that the temporal distance between neighboring pairs of packets of two adjacent networks shrinks (or grows) as time proceeds, and the amount of shrinkage per interval is a constant value $\gamma$. The parameter $\gamma$ depends on the interval lengths of a pair of networks examined. Only if this distance reaches $0$, the offset needs to be re-computed and the new offset shrinks with multiples of $\gamma$ again. We present a theory on how to compute $\gamma$ and, based thereupon, on which events can be skipped without affecting the simulation results. Using the value of $\gamma$, we systematically determine and skip the pairs of succeeding events whose temporal distances are too large to collide. 

While we present adaptive event skipping in the context of collision simulations for BLE networks, our proposed technique is generic and can be used for accelerating discrete event simulation of different problems in which overlaps of cyclic events are of interest. Besides simulating collisions of other protocols than BLE (e.g. ANT/ANT+), the technique can also be used for studying e.g. schedules of multiple periodic software tasks with different periods or colliding requests for a shared resource.

In this paper, we make the following contributions compared to the literature:
\begin{asparaitem}
\item We present a novel simulation speedup technique, which can reduce the simulation time by up to 2 orders of magnitude. It is applicable to all simulations in which possible interactions between multiple, purely-cyclic events need to be studied.
\item We evaluate the resulting speedup of our proposed technique by running a large number of simulations. 
\end{asparaitem}
\vspace{0.05cm}
The rest of this paper is organized as follows. In Section \ref{sec:related_work}, we present related work. Next, in Section \ref{sec:simulation_model}, we briefly introduce the relevant parts of the BLE protocol and describe a valid simulation model for it in detail. Our proposed speedup technique is presented in Section \ref{sec:speedup} and evaluated in Section \ref{sec:evaluation}. Finally, we conclude our results in Section \ref{sec:concluding_remarks}.

\section{Related work}
\label{sec:related_work}
\subsection{Collision Modeling}
Analytical collision modeling proved to be effective in analyzing the performance and energy consumption of wireless sensor networks. 
For example, a probabilistic analysis of interference and collisions of power-aware ad-hoc networks using the 802.11 MAC protocol have been proposed in~\cite{Gobriel:INFOCOM04}.
A Markov chain-based model for 802.11 has been presented in~\cite{wu:infocom02}. However, the protocols considered are not of cyclic nature, and therefore these models cannot be applied to the Bluetooth Low Energy or ANT/ANT+ protocol. 
For BLE, to the best of our knowledge, the only known analytical model \cite{liu:13} is based on the classical ALOHA analysis.
However, this model is only applicable to the \textit{advertising mode} of BLE (in contrast to the \textit{connected mode} considered in this paper), which is not purely cyclic. Further, it is restricted to all interfering networks having the same parametrization. 

In the connected mode of BLE, in which packets are exchanged in a purely cyclic manner, one can assess the collision probability of two networks with connection intervals $T_{c,1} < T_{c,2}$ by determining the fraction of time used for packet transmission and the shorter interval, $T_{c,1}$. Given a packet length of $d$ for both networks, the collision probability for a packet of network $2$ is $\frac{2 \cdot d}{T_{c,1}}$. 
However, as soon as the number of interfering networks is increased beyond 2, the amount of time the channel is busy is hard to determine, since packets from multiple interferers might overlap with each other. Moreover, the time-intervals in which the channel is vacant might become shorter than one packet length $d$, and therefore any packet starting within such an interval also collides. Analytical models accounting for this are only known for certain distributions of packet arrival times, e.g., the exponential distribution assumed in \cite{liu:13}. However, for purely cyclic protocols, to the best of our knowledge, none of the known distributions applies.  

Further, even in the few cases for which analytical models exist, different transmit power levels and different distances among the nodes are not accounted for by these models. 
In typical MANETs, it is common for a transmitting node to change its location over time.
Discrete event simulations are inevitable to overcome these challenges. In the following, we give a brief overview on previous work related to such simulations.

Our work is also related to \cite{kindt:15c}, where the neighbor discovery latency of cyclic protocols has been studied.
%
%

\subsection{Wireless Protocol Simulation}
Simulating wireless networks is a widely-used technique for studying the behavior of wireless networks, such as collision rates or packet latencies. For example, NS2 \cite{ns:15}, Omnet++ \cite{omnet:15} and TOSSIM \cite{tossim:03} are widely used. Most of them, e.g., Avrora \cite{Titzer:ipsn05}, rely on next-event time advance schemes same as the one used in this paper.
However, these simulation environments do not offer support for the BLE protocol yet, as the BLE protocol has been introduced only recently.
Some simulation environments to estimate certain aspects of BLE have been proposed. These simulators are mostly limited to the problems studied in these papers. For example, \cite{Mikhaylov:pimrc14} uses a dedicated discrete event simulator to simulate the throughput and collision properties during the connection setup. \cite{kalaa:14} has simulated the BLE channel hopping algorithm using another dedicated simulator. Other work has simulated BLE transceiver hardware, e.g., in \cite{li:13}. An algorithm to simulate the neighbor discovery procedure of BLE has been presented and released into public in \cite{kindt:corr14}. However, this algorithm can neither be regarded as a full-featured protocol simulator which is capable of estimating collision rates, nor contains any speedup techniques comparable to this paper.

Since there is no standard simulation environment for BLE, we developed our own custom simulator for this paper. Its architecture and simulation model are described in the next section.

\section{Simulation Model}
\label{sec:simulation_model}
In this section, we first briefly introduce the relevant parts of the BLE protocol. Next, we describe the simulation model for BLE used in our proposed simulator.

\subsection{Bluetooth Low Energy}
\vspace*{-0.2cm}
\subsubsection{Over-the-Air Packet Flow}
\begin{figure}[htb]
\centering
\includegraphics[width=\linewidth]{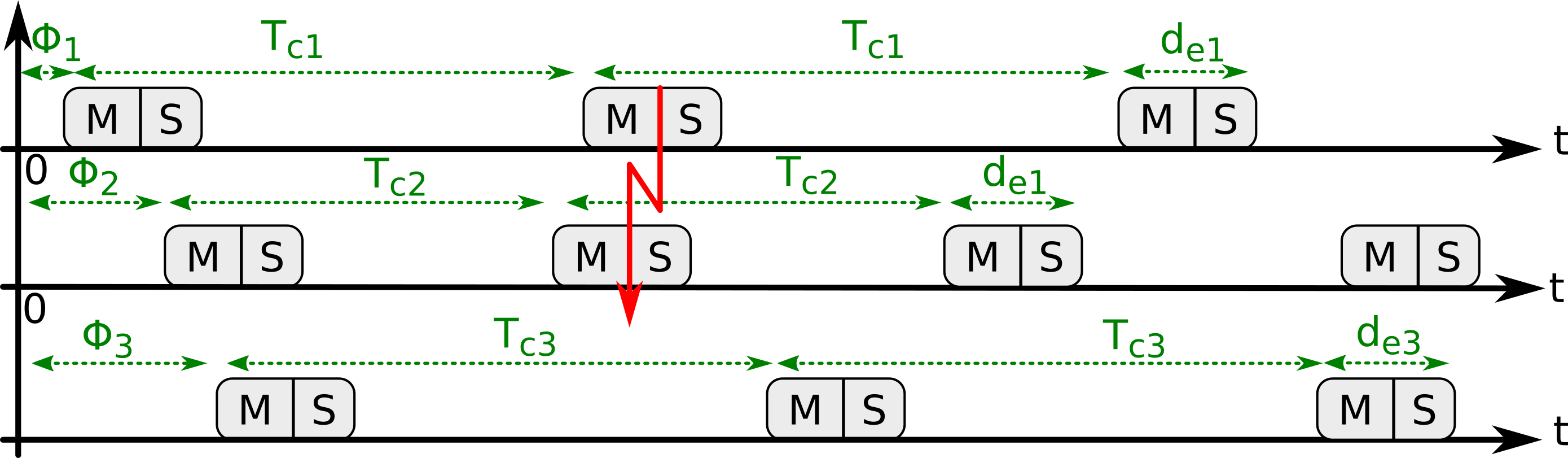}
\caption{Packet transmission scheme of 3 interfering networks.}
\label{fig:ble_collision_situation} 
\end{figure}
We briefly introduce the aspects of BLE which are relevant for the collision simulation. The descriptions are based on the official Bluetooth specification \cite{bleSpec4.2}. For the sake of exposition, we assume that each network consists of one master and one slave without loss of generality. Our speedup technique can also be applied to networks with multiple slaves. In BLE, the \textit{connected mode} is usually applied for exchanging packets, which works as follows. During the connection setup, the master and the slave have negotiated a \emph{connection interval}, $T_c \in [7.5~ms, 10.24~s]$, which is a multiple of $1.25~ms$. Packets are exchanged as shown in Figure~\ref{fig:ble_collision_situation} for 3 different networks, once the connection has been established. A connection event takes place after each $T_c$ amounts of time.

At each connection event, the master sends one packet to the slave (indicated by \textit{M} in Figure~\ref{fig:ble_collision_situation}), whereas the slave can either respond with another packet (indicated by \textit{S}) or remain asleep if there is no data to send. Between two packets, there is a short duration of idleness called the \textit{interframe space}, which is not shown in the figure.
In this paper, we assume that there is always one pair of packets per connection event. A collision occurs if one of the packets is sent at a point in time at which the transmission of a different network takes place, as shown for the second packets of networks 1 and 2 in Figure~\ref{fig:ble_collision_situation}. Since all networks operate independently of each other, they have different offsets $\Phi_n$ to an (arbitrary) origin of the time axis. The collision rate for any of these networks is determined by the values of the initial offsets $(\Phi_1, \Phi_2, \Phi_3)$, the connection intervals $T_{c,n}$ and by the durations $d_{e,n}$ of the connection events. While these durations might actually vary within one network, for the sake of exposition, we assume that they remain constant within every network. For a given number of bytes, the duration of a connection event is as follows:
\begin{equation}
\label{eq:eventDurationBytes}
d_{e} = (n_{pkg,m} + n_{pkg,s}) \cdot 8 \cdot 10^{-6} \cdot d_{IFS}.
\end{equation}
In Equation \eqref{eq:eventDurationBytes}, $n_{pkg,m}$ and $n_{pkg,s}$ is the number of bytes per packet of the master or the slave, respectively. The factor $8 \cdot 10^{-6}$ results from the over-the-air symbol rate of BLE, which is defined as $1~MHz$. 

To avoid ambiguities, we distinguish between BLE connection events and simulation events (discrete events executed by the simulator) by always writing explicitly \textit{connection event} when referring to the exchange of BLE packet pairs. 

\subsubsection{Channel Hopping Mechanism}
To reduce the risk of collisions, both with neighboring BLE networks (as considered in this paper) and also with other communication standards, BLE makes use of a channel-hopping mechanism. One among 37 channels is selected for each connection event. Under the assumption that all channels are used, BLE increments the index of the next channel to use $\tau_h$ by a certain value and divides it modulo 37.

\subsection{Simulation Infrastructure}
For demonstrating and evaluating our proposed speed\-up technique, we have implemented our own simulator. We did not use any of the existing, publicly available simulators such as NS2 \cite{ns:15} or OMNet++ \cite{omnet:15} for the following reasons.
\begin{asparaenum}
\item To the best of our knowledge, none of them comes with a full implementation of BLE.
\item The existing solutions potentially contain features which are not necessarily needed for the collision rate estimation.
In contrast, our custom simulator is a minimal setup for simulating BLE packet collisions. This allows for an evaluation of the speedup in a more comparable and fair manner.
\end{asparaenum}
\vspace{0.05cm}
\begin{figure}[htb]
\centering
\includegraphics[width=\linewidth]{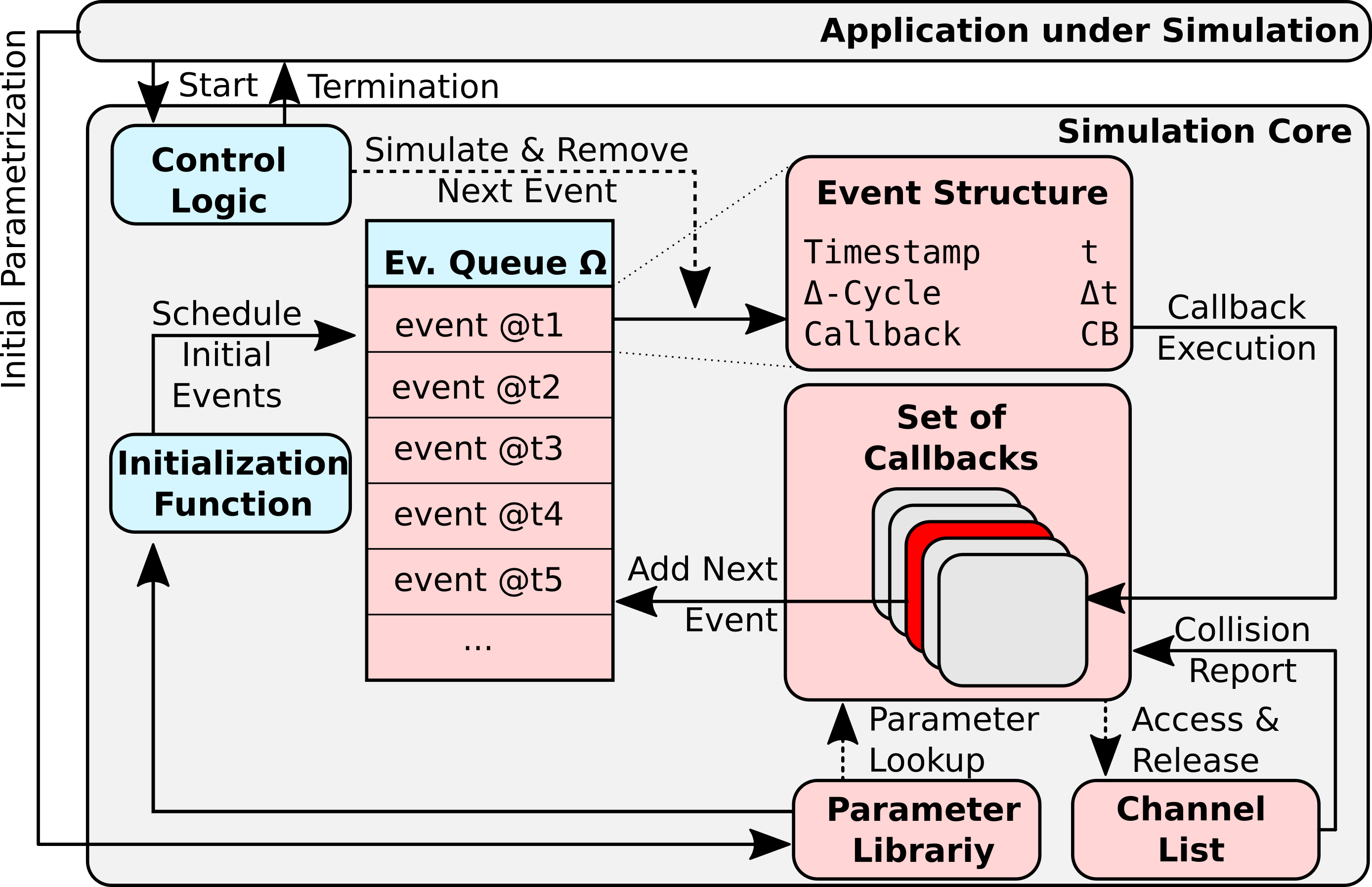}
\caption{Architecture of the simulator.}
\label{fig:simulation_architecture} 
\end{figure}
Our simulator is a next-event time advance-based system written in C++, which implements the state of the art known from the literature \cite{law:13} \cite{mansharamani:97}. Its architecture is shown in Figure~\ref{fig:simulation_architecture}. The simulation core is built around the \textit{event queue} $\Omega$. It is implemented as a priority queue which consists of a list of pointers to \textit{event structures}. This data structure contains all data which is necessary to represent a simulation event. 
The timestamp $t$ and the delta-cycle $\Delta t$ determine the point in simulated time the event is executed at.
If $t$ is equal for two events, then the event with the higher delta-cycle is simulated first.
In addition, a pointer $CB$ to an event-callback function, which is executed once the event is simulated, is stored. 
At the beginning of a simulation, the \textit{initialization function} generates an initial set of events (viz, the first event of each network). The simulation \textit{control logic} always removes the topmost element from the queue and executes the event callback function which is assigned to the event. A collection of callback functions contains the necessary procedures for simulating BLE connection events and for detecting channel collisions, as described below. Further, these callbacks can generate future events which are inserted into the priority queue such that they are sorted in ascending order of their timestamps. A \textit{parameter library} contains all data for each network $n$, such as the initial offsets $\Phi_n$ and the connection intervals $T_{c,n}$. Each  callback function can access this data to schedule the next events. In addition, a \textit{channel list} keeps track of all channels for detecting collisions. In particular, an event can access or release any of the 37 channels. 
The simulation is terminated by the control logic once a certain time-limit has been reached. Time is counted in microseconds, which is a sufficient resolution to account for all events of the BLE protocol without any deterioration of the simulation results. An \textit{application under simulation} configures the parameter library with the appropriate values (e.g., connection intervals, packet lengths) to represent a certain scenario and then runs the simulation until it terminates. This is repeated multiple times with different initial offsets $\Phi$ and connection intervals $T_c$ to obtain data on the collision rate. The collision rate is defined by the number of packets collided divided by the number of packets sent in a certain network of interest (NoI).

\subsection{Simulating BLE}
\label{sec:simulationModel}
\subsubsection{Mapping Collision Events to Simulation Events}
For each BLE connection event, a series of simulation events is generated. 
Each connection event consists of two packets separated by the interframe space.
For each of these packets, a check for collisions is conducted at the beginning and at the end of the transmission. The detection of collisions is guaranteed whenever packets of at least two different networks overlap.
Therefore, 2 simulation events are needed at the beginning and another 2 at the end of each packet: A separate pair of events to access/release the channel and to check for collisions.

In our simulator, the first simulation event in each connection interval adds the first simulation event for the next interval to the event queue. In addition, each event adds its direct successor of the same connection interval to the queue.
\subsubsection{Simulation Termination}
The simulation is terminated after a predefined duration of $d_{sim}$ time-units.
In fact, there is an optimal duration for each given scenario. We consider $N$ networks with a set of connection intervals $\vec{T_c}$ and a set of initial offsets $\vec{\Omega}$ with $\Omega_n \leq T_{c,n}\mbox{  }\forall n \in \vec{\Omega}$. After a certain simulation duration, the offsets between all neighboring packets of all networks will be identical to the initial ones.
Simulation beyond it would yield the same results repeatedly.
This duration is the optimal simulation duration, and hence $d_{sim}$ needs to be computed as follows.
\begin{equation}
\label{eq:dsim}
d_{sim} = LCM(\sigma \cdot T_{c,0}, \sigma \cdot T_{c,1},...,\sigma \cdot T_{c,n}) + 2 \cdot d_{pkg} + d_{IFS}
\end{equation}
with $\sigma = LCM(37, \tau_h)$. 
If $d_{sim}$ is selected according to Equation \eqref{eq:dsim}, the simulated collision rate is the exact result for a given parametrization $(\vec{T_c}$, $\vec{\Omega})$.
Such simulations can become very complex and therefore time-consuming, which will be addressed with a speedup technique in the next section.

\section{Adaptive Event Skipping}
\label{sec:speedup}
\vspace*{-0.2cm}
\subsection{Overview}

\begin{figure*}[htb]
\begin{center}
\includegraphics[width=1.0\linewidth]{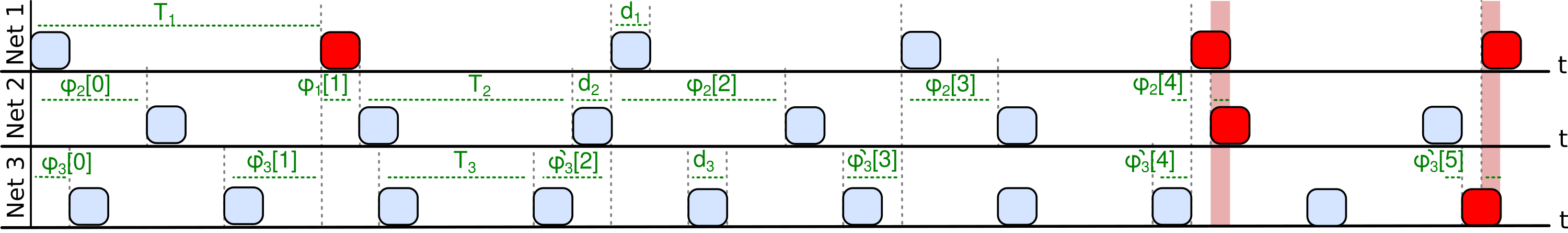}
\vspace*{-0.7cm}
\caption{Skipping events.\vspace*{-0.8cm}}
\label{fig:skipManager_scheme}
\end{center}
\end{figure*}

In this section, we present a novel technique to predict the events that have a chance of colliding, and skipping all events for which a collision can be excluded. An additional software module which we call the \textit{skip manager} is added to the simulation infrastructure depicted in Figure \ref{fig:simulation_architecture}. Whenever the first callback of a connection event is executed, this software module is run to predict the next point in time an event has to be scheduled at for the current network. All events in between can be skipped, leading to a significant reduction of events that need to be simulated. 
It is guaranteed that all events which are skipped do not collide, and therefore the resulting collision rate would still be the same.
Before we go into the details, we make the following assumptions:
\begin{asparaenum}
\item By a discrete event simulation of $N$ different networks, the collision rate of one network $n_{oI}$ which is referred to as the \textit{network of interest} shall be assessed. We assume that each network consists of one master and one slave which exchange one pair of packets per connection event. However, our proposed skipping scheme is not restricted to this scenario, and can handle networks with arbitrary numbers of slaves and pairs of packets.
\item We assume $T_{c,n+1} <= T_{c,n}$, without loss of generality.
\item The first event of each network $n \ne n_{oI}$ starts with an offset $\Phi_{n}[0]$ relative to an (arbitrary) origin of the coordinate system. We define $\Phi_{NoI}[0] = 2 \cdot d_{pkg} + d_{IFS}$ and further assume that $\Phi_{n}[0] \le T_{c,n}, \forall n$. This means that the relative offsets $\varphi_n[0]$ between the first event of network $n$ and the first event of the NoI are uniformly distributed.
\end{asparaenum}
\subsection{Skipping Events}
\label{sec:smOverview}

The skip manager predicts the next point in time at which another connection event needs to be scheduled rather than always adding a new event for the succeeding connection interval.
Simulation events of the NoI are inserted to the event queue by the event callbacks for the remaining networks, rather than by instances called for the NoI. It is realized in this way because the collision prediction by the skip manager is done in a pairwise manner for each pair of $n$ and the NoI.

Consider the situation depicted in Figure \ref{fig:skipManager_scheme}, where 3 networks periodically exchange packets with their own intervals. Each connection event is depicted as a rectangle of length $d_n$ and colliding packets are highlighted. The duration $d_n$ is defined as $d_n = 2 \cdot d_{pkg} + d_{IFS}$.
The NoI is network 1 in this example.
We define $\varphi_n[k]$ as the distance between the beginning of event $k$ of the NoI and the beginning of the left neighbor event (temporally closest event to the left in the timeline) of network $n$. Accordingly, $\varphi'_n[k]$ is the distance between the NoI-event $k$ and the right neighbor event (defined vice versa) of network $n$. 
With increasing numbers $k$, the value of $\varphi_n[k]$ (or $\varphi'_n[k]$, respectively) grows or shrinks by a constant amount of time. For example, consider network 2. Here, $\varphi_2[2] > \varphi_2[3] > \varphi_2[4]$ with $\varphi_2[2] - \varphi_2[3] = \varphi_1[3] - \varphi_2[4] = \gamma_2$. Similarly, for network 3, it is $\varphi'_3[1] > \varphi'_3[2] > \varphi'_3[3] > \varphi'_3[4]$ with $\varphi_3'[1] - \varphi_3'[2] = \varphi_3'[3] - \varphi_3'[2] = \varphi_3'[4] - \varphi_3'[3] = \gamma_3$. The growth or shrinkage per event is constant as long as the event of reference does not change. The first two connection events of network $2$ exemplify a situation in which the event of reference changes without a collision taking place. Between $\varphi_2[1]$ and $\varphi_2[2]$, one event of network 2 needs to be skipped because $\varphi_2[1] - \gamma_2 < 0$, and hence the resulting event would become the left neighbor instead of the right neighbor, as required by the definition of $\varphi$. We denote situations in which $\varphi_n[k]$ shrinks with multiples of $\gamma_n$ as \textit{shrinking processes}, whereas situations in which $\varphi_n[k]$ grows are denoted as \textit{growing processes}.

Given any offset $\varphi_n[k]$ and $\gamma_n$, it is possible to predict the next point in time at which the events of both networks either overlap or the element of reference changes. For example, in network 2, a necessary condition for a collision is $\varphi_2[k] < d_1$. Given any offset $\varphi_2 > d_1$, the number of events $k_1$ of network 1 which can be skipped safely regarding collisions with network 2 is 
\begin{equation}
k_1 = \left\lfloor \frac{\varphi_2 - d_s}{\gamma_2} \right\rfloor.
\end{equation}

This concept can be extend towards all cases (viz, all connection intervals and event durations).
In the following, we first describe in detail how the simulation infrastructure makes use of the skip manager.

\subsection{Integration of the Skip Manager}
The skip manager is executed in the first event callback of all networks except the NoI.
For each network $n$, the skip manager determines a vector $\vv{k} = [k_{NoI}, k_n]$, which is the number of connection intervals after which the next connection events of network $n$ and the NoI need to be scheduled.
The prediction carried out for each network is based on two connection intervals $T_l$ and $T_h$ defined as follows.
\begin{equation}
\begin{array}{lr}
T_l = min(T_{c,n}, T_{c,NoI}), &
T_h = max(T_{c,n}, T_{c,NoI}). \\
\end{array}
\end{equation}
We distinguish between the network \textit{H} which has the higher connection interval $T_h$ and the network $L$ which has the lower connection interval $T_l$. 
From these intervals, the offset between a connection event $l'$ of network $n$ and the corresponding event $l$ of the NoI is computed as follows.

\begin{equation}
\varphi = \left\{\begin{array}{ccc} t_n[l'] - t_{NoI}[l]& \mbox{, if } & T_{NoI} > T_n, \\t_{NoI}[l] - t_n[l'] & \mbox{, else.} &\end{array} \right.
\end{equation}
$t_n[l']$ is the timestamp of the simulation event $l'$ of network $n$. $t_{NoI}[l]$ is the timestamp of the related event $l$ of the NoI. To determine which NoI-event $l$ is related to event $l'$ of network $n$, an array $\vv{t_{NoI}}$ contains the time-instances of the previously predicted NoI-events for each network $n \neq n_{NoI}$. All of its elements are initialized with the timestamp of the first connection event of the NoI. As soon as a connection event of network $n$ has been simulated, $t_{NoI}[n]$ is set to the point in time predicted by the skip manager:
\begin{equation}
\label{eq:tNoi}
t_{NoI}[n] = t_{NoI}[n] + k_{NoI} \cdot T_{c,NoI}.
\end{equation}

After each prediction, the event for the NoI is scheduled at $t_{NoI}[n]$ according to Equation \eqref{eq:tNoi}, and the event for network $n$ at $k_n \cdot T_n$ time-units in the future.
Now, we present steps to calculate $\gamma$ parameter, which is required to calculate $k_l$ and $k_h$ values.


\subsection{Computing the Value of $\gamma$}
As we briefly described in Section~\ref{sec:smOverview}, $\gamma$ is the amount of offset shrinkage or growth per event of the larger interval. 
$\gamma$ is a function of $T_l$ and $T_h$, defined as the difference of the larger interval from appropriate multiples of the smaller one.
For a shrinking process, it is $\gamma  = T_{h} - \lfloor \frac{T_{h}}{T_{l}}\rfloor \cdot T_{l}$ and for growing processes, it is $\gamma  = \lceil \frac{T_{h}}{T_{l}}\rceil \cdot T_{l} - T_{h}$ \cite{kindt:15c}.
To determine the mode $m \in (s,g)$, which indicates whether the process is shrinking($s$) or growing($g$), Equation \eqref{eq:processMode} can be used.
\begin{equation}
\label{eq:processMode}
m = \left\{\begin{array}{ccc} s & \mbox{, if } &\left\lceil \frac{T_{h}}{T_{l}}\right\rceil \cdot T_{l} - T_{h} > \frac{1}{2} T_l, \\g & \mbox{, else.} &\end{array} \right.
\end{equation}

\subsection{Prediction Scheme}
The core of the skip manager is the event predictor. Given $T_l$, $T_h$, $d_l$, $d_h$ and $\varphi$, its purpose is finding values for $k_l$ and $k_h$ which determine the number of intervals of the network with the shorter ($k_l$) and larger ($k_h$) connection interval that can be skipped, respectively.
For the sake of illustration, we assume $d_l = d_h = d$. 
Depending on the input parameters, there are multiple cases which need to be taken into account.
\subsubsection{Constant Situations: $\gamma=0$}
If $\gamma = 0$, the offset $\varphi$ between two neighboring events always remains constant. Hence, if the current pair of connection events overlaps, all future events of the network with the larger interval will overlap as well. In contrast, if the current pair of events does not overlap, all future pairs of events cannot overlap either. However, the current pair of events that is examined might not be the closest neighbors and hence multiple events of network L need to be examined.

If $\varphi < - d_s$, the event of network L takes place before the one of network H. In this case, we compute $k_s$, which is the number of intervals of length $T_l$ which fit into $|\varphi|$. We then correct the offset to the one of the nearest neighbor, if applicable. It is:
\begin{equation}
\begin{array}{lr}
k_s = \left\lfloor \frac{-\varphi}{T_l} \right\rfloor, & \varphi = \varphi + k_s T_h.\\
\end{array}
\end{equation}

We can now check whether the current event (with the corrections of $\varphi$) matches. Therefore, if $-d \leq \varphi \leq d$, it is:
\begin{equation}
\begin{array}{lr}
k_h = 1, & k_l = \left\lfloor \frac{T_h}{T_l} \right\rfloor + k_s.
\end{array}
\end{equation}
Otherwise, we consider cases in which the currently examined event of network L does not collide with the current event of network H, but its successor in the next connection interval of network H does. Therefore, a second offset $\varphi_r$ needs to be defined as follows.
\begin{equation}
\begin{array}{lr}
\varphi_r = \varphi + k_r \cdot T_l, & k_r = \left\lceil \frac{T_h - \varphi - d_s}{T_l} \right\rceil.\\
\end{array}
\end{equation}
The collision in the next interval of network H occurs if $T_h  - d < \varphi_r < T_h + d$. Then it is $k_h = 1$ and
\begin{equation}
\begin{array}{l}
k_l =  \left\{\begin{array}{ccc} \left\lceil \frac{T_h - \varphi - d_s}{T_l} \right\rceil & \mbox{, if } & k_r  = 0, \\k_r & \mbox{, else.} &\end{array} \right.
\end{array}
\end{equation}
Finally, if the two networks never collide, we set $k_l \gets \infty$ and $k_r \gets \infty$, thereby indicating that no further events need to be scheduled for the currently examined event to assess the collision rate with the NoI.

\subsubsection{Growing Situations: $T_{h} - \left\lfloor \frac{T_{h}}{T_{l}}\right\rfloor \cdot T_{l} > \frac{1}{2} \cdot T_l$}
In growing situations, a match occurs if the offset $\varphi$ approaches $1 \cdot T_h - d$. This could either be the case for the currently examined pair of connection events, or for future ones.  Hence, multiple offset corrections have to be carried out and different subcases exist.
\begin{figure}[bt]
\begin{center}
\includegraphics[width=1.0\linewidth]{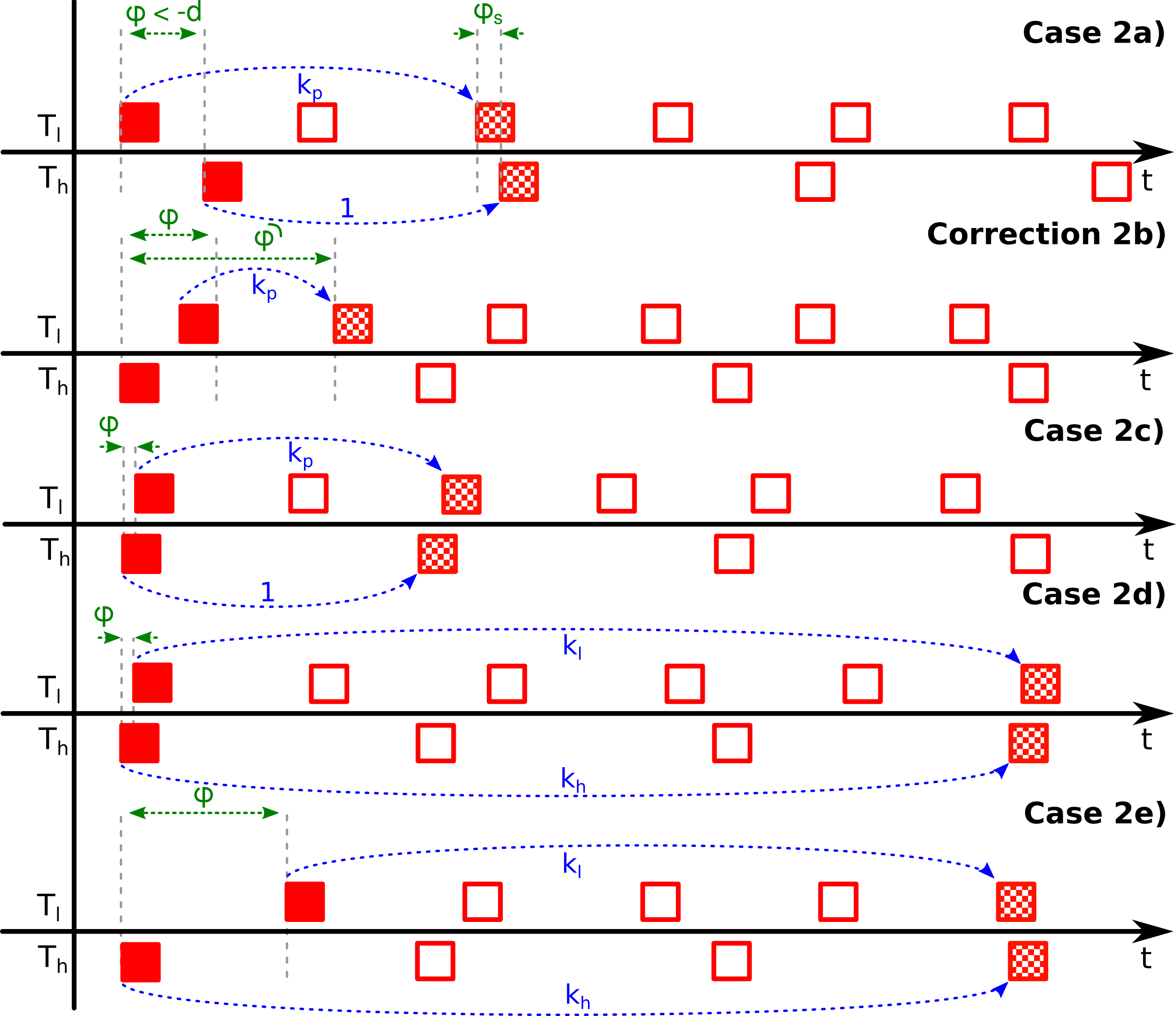}
\caption{Different subcases depending on the initial offset $\varphi$ in growing situations (case 2).}
\label{fig:case2}
\end{center}
\end{figure}
Figure \ref{fig:case2} depicts all possible subcases. The colored boxes depict the input events of the prediction. The hatched boxes depict the predicted events, and the hollow ones depict the skipped ones.
For the case 2a), we consider negative offsets $\varphi < 0$. This means that the connection event of network L takes place before the connection event of network H. We introduce a correction factor $k_s$ which represents the number of intervals of length $T_l$ that can occur without reaching the next event (i.e., the event being temporally on the right) of network H as follows.

\begin{equation}
\label{eq:case2kp}
k_p = \left\lfloor \frac{ T_h - \varphi+ d_s}{T_l}\right\rfloor.
\end{equation}
The offset between this event of network L and the next event of network H is defined as $\varphi_s = T_h - \varphi - k_p T_l$. Depending on the values of $\varphi$ and $\varphi_s$, multiple subcases exist. \\
\textbf{Case 2a): $\mathbf{\boldsymbol{\varphi < -d \land -d \leq \varphi_s \leq d}}$:}
In this case, the next event of network H matches with an event of network L. We set
$k_l = k_s$ and $k_h = 1$. Otherwise, we apply correction 2b) to get a corrected offset and continue checking for the remaining cases with the corrected offset. In addition, correction 2b) is also applied for offsets $\varphi > -d_s$, since cases 2c) - e) require that $T_h - \varphi - d < T_l$.\\
\textbf{Correction 2b): $\boldsymbol{\mathbf{\varphi_s \notin [-d, d] \lor \varphi \geq -d}}$:}\\
In this case, the offset $\varphi$ is corrected by $k_p$ intervals of $T_l$, such that the reverse offset $\varphi_s$ becomes smaller than $T_l$. This means that collisions between the next connection event of network H and its leftmost neighbor of network L are to be examined in the remaining cases. 
The effective offset is $\varphi' = \varphi + k_p \cdot T_l$. We set $\varphi \gets \varphi'$ and examine the corrected offset for the further cases c) - e).\\

\noindent\textbf{Case 2c): $\mathbf{\boldsymbol{-d \leq \varphi\leq d \land \varphi+ \gamma < d}}$:}
In this case, the currently examined connection events match and the next connection event of network H matches one of network L as well. Therefore, we set
\begin{equation}
\begin{array}{lr}
k_h = 1, & k_l = \left\lceil \frac{T_h - \varphi- d}{T_l}\right\rceil + k_p. \\
\end{array}
\end{equation}
\textbf{Case 2d): $\mathbf{\boldsymbol{-d \leq \varphi\leq d \land \varphi+ \gamma \geq d}}$:}
In this case, the currently examined pair of connection events matches, but the next one of network \textit{H} does not overlap with any connection event of network \textit{L}. The number of intervals $T_h$ to skip is determined by the number of $\gamma$-intervals that lie within the distance between the next connection event of network H and its rightmost left neighbor of network L. With $k_{pp} = \lfloor \frac{T_h - \varphi}{T_l} \rfloor$, it is:
\begin{equation}
\begin{array}{lr}
k_h = \left\lceil \frac{T_h - \varphi- k_{pp} T_l - d}{\gamma}\right\rceil + 1, & k_l = \left\lceil \frac{k_h T_h - \varphi- d}{T_l}\right\rceil + k_p,\\
\end{array}
\end{equation}
\textbf{Case 2e): $\boldsymbol{\mathbf{\varphi\notin [-d, d]}}$:}
For all other cases, the current pair of events does not match. We determine the distance until the next match occurs by computing the number of $\gamma$-intervals which fit within the remaining distance of the next event of network H and its rightmost left neighbor of network L. $k_l$ and $k_h$ are computed as follows.
\begin{equation}
\begin{array}{lr}
k_h = \left\lceil \frac{T_h - \varphi- d}{\gamma}\right\rceil + 1, & k_l = \left\lceil \frac{k_h T_h - \varphi- d}{T_l}\right\rceil + k_p. \\
\end{array}
\end{equation}

\subsubsection{Shrinking Situations: $T_{h} - \left\lfloor \frac{T_{h}}{T_{l}}\right\rfloor \cdot T_{l} < \frac{1}{2} \cdot T_l$}
\begin{figure}[bt]
\begin{center}
\includegraphics[width=1.0\linewidth]{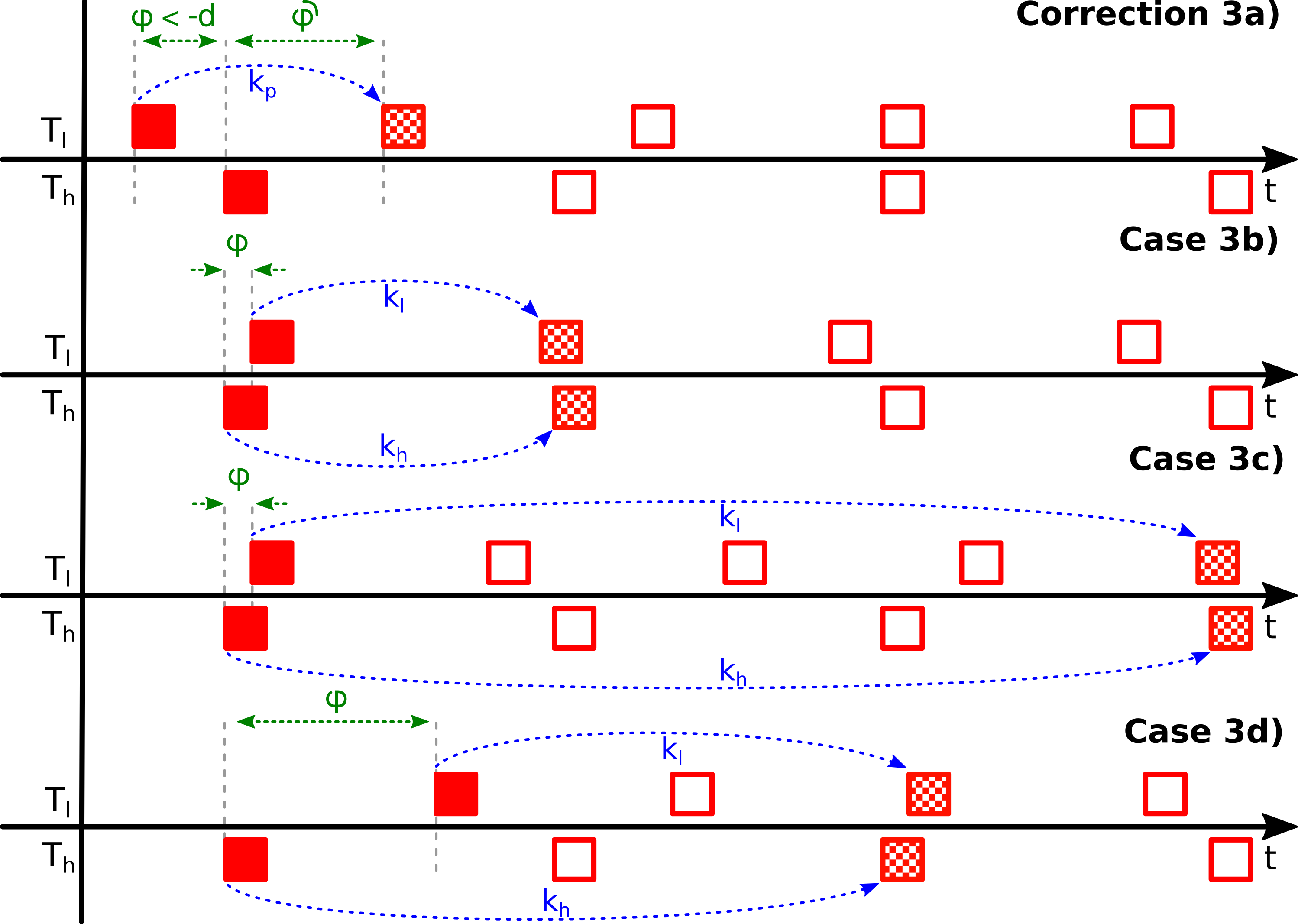}
\caption{Different subcases depending on the initial offset $\varphi$ in shrinking situations (case 3).}
\label{fig:case3}
\end{center}
\end{figure}

The subcases for shrinking situations are shown in Figure \ref{fig:case3}.
First, largely negative offsets are corrected.\\
\textbf{Correction 3a): $\boldsymbol{\mathbf{\varphi< -d}}$:}
If the initial offset has a negative value below $-d$, the event of network \textit{L} examined cannot shrink towards the event of network \textit{H} since it is already temporally left of it. Therefore, we consider the next event of network \textit{L} which is temporally right of the one considered for network \textit{H}.
We define $k_p$ as $\lceil \frac{-\varphi}{T_l} \rceil$.
If $\varphi\geq d$, $k_p$ is set to $0$. The corrected offset $\varphi'$ is $\varphi+ k_p \cdot T_l.$
We set $\varphi\gets \varphi'$ and continue checking for the different cases with the corrected offset.\\
\textbf{Case 3b): $\boldsymbol{\mathbf{(-d \leq \varphi\leq d) \land (\varphi- \gamma \geq -d)}}$:}
In this case, both the current and the next event of network \textit{H} overlap with an event of network \textit{L}. It is:
\begin{equation}
\begin{array}{lr}
k_h = 1, & k_l = \left\lfloor \frac{T_h}{T_l} \right\rfloor + k_p.
\end{array}
\end{equation}
 \textbf{Case 3c): $\boldsymbol{\mathbf{(-d \leq \varphi\leq d) \land (\varphi- \gamma < -d)}}$:}
Next, we consider the case that the current and of network H overlaps with any event of network L, but its successor does not. $k_l$ and $k_h$ are defined as follows.
\begin{equation}
\begin{array}{lr}
k_h = \left\lceil \frac{T_l +\varphi- d}{\gamma} \right\rceil, & k_l = \left\lceil \frac{k_h T_h - \varphi- d}{T_l} \right\rceil + k_p.
\end{array}
\end{equation}
 \textbf{Case 3d): $\boldsymbol{\mathbf{\varphi\notin [-d, d]}}$:}
In this case, the currently examined pair of events does not match. We shrink the distance $\varphi$ between them below $d_s$, until an event of network H overlaps with network L or the connection event of reference changes without a collision. We therefore compute $k_l$ and $k_h$ as follows.
\begin{equation}
\begin{array}{lr}
k_h = \left\lceil \frac{\varphi- d}{\gamma} \right\rceil, & k_l = \left\lceil \frac{k_h T_h - \varphi- d}{T_l} \right\rceil + k_p.
\end{array}
\end{equation} 

\subsection{Estimating the Number of Packets Sent}
Even though the events without possibility of collision are skipped, we still need to compute the number of packets sent to calculate the collision probability.
An easy-to-use solution is to compute them based on the simulation duration.
Given a total simulated time $d_{sim}$, $N_{total,i}$ of network $i$ can be computed according to Equation \eqref{eq:nTotal}.
\begin{equation}
\label{eq:nTotal}
N_{total,i} = \left\{\begin{array}{clr} 1 & \mbox{, if} & d_{sim} = 0 \\ \left\lceil \frac{d_{sim}}{T_{c,i}}\right\rceil &\mbox{, else.}  & \end{array}\right.
\end{equation}

In the next section, we evaluate the amount of speedup obtained in different scenarios.

\section{Evaluation}
\label{sec:evaluation}
By applying the proposed adaptive event skipping, significantly less number of events need to be simulated. In addition, the number of events which need to be sorted, inserted into and removed from the priority queue $\Omega$ is decreased. 
On the other hand, predicting a number of connection intervals to be skipped introduces some computational overhead within the first callback function of each connection event, which has to be correctly evaluated.
This section evaluates the achievable speedup in realistic situations. We determine the minimum speedup (which might also be slightly negative due to the overhead), the mean and the maximum speedup in a large number of simulated situations. Thereby, we show that for most of them, a significant speedup is achieved, whereas the penalty in the rare cases in which the simulation is decelerated is negligible. 

\subsection{Evaluation Methology}
In what follows, we conduct a series of experiments with a large number of randomly generated situations.
We consider a fixed number of networks $N$. Each of them choses its connection interval $T_{c,n}$ randomly between $T_{min}$ and $T_{max}[k]$. $T_{min}$ is fixed while $T_{max}[k]$ depends on an index $k$. For each value of $T_{max}[k]$, we repeat every experiment with random connection intervals and offsets $N_{r}$ times. $T_{max}[0]$ is initialized by $T_{min}$ and $T_{max}[k]$ is increased by $1.25~ms$ for every $k$ after each $N_{r}$ simulation runs. All networks $n \neq n_{oI}$ have a random, uniformly distributed starting offset between $0$ and $T_{c,n}$. $T_{min}$ is set to $7.5~ms$, the minimum connection interval of BLE.
Among the $N_{r}$ repetitions for each value of $T_{max}[k]$, we determine the minimum, mean and maximum speedup in terms of the number of events processed per simulation and in terms of CPU-time. The CPU-time is determined by the \texttt{clock()}-function of the GNU C library \cite{gnu:15}.
The timing measurement does not account for the initialization phase of the simulation, but for the more relevant processing phase.
We chose $d_{pkg} = 296~\mu s$, for both the master and the slave, which corresponds to a packet length of $37~Bytes$ including all overheads.
The hop-increment $\tau_h$ of the channel-hopping algorithm is set to 1.

The resulting numbers of collisions and packets sent are cross-checked with a simulation run without the proposed scheme to ensure the correctness. 
Experiment 1 is conducted on an Intel Core i7-4600U processor, whereas the remaining ones are run on an AMD Athlon II X2 250e processor. Only one core is used simultaneously by each experiment.
Some experiments have become excessively long and were aborted after $\SI{30}{h}$ of wall-clock time.
The sum of simulation time is given for each experiment.

\subsection{Evaluation Results}
\subsubsection{Collision Rate and Speedup}
\begin{figure*}[tbh]
\begin{center}
\begin{subfigure}[b]{0.49\textwidth}
\includegraphics[width=1.0\linewidth]{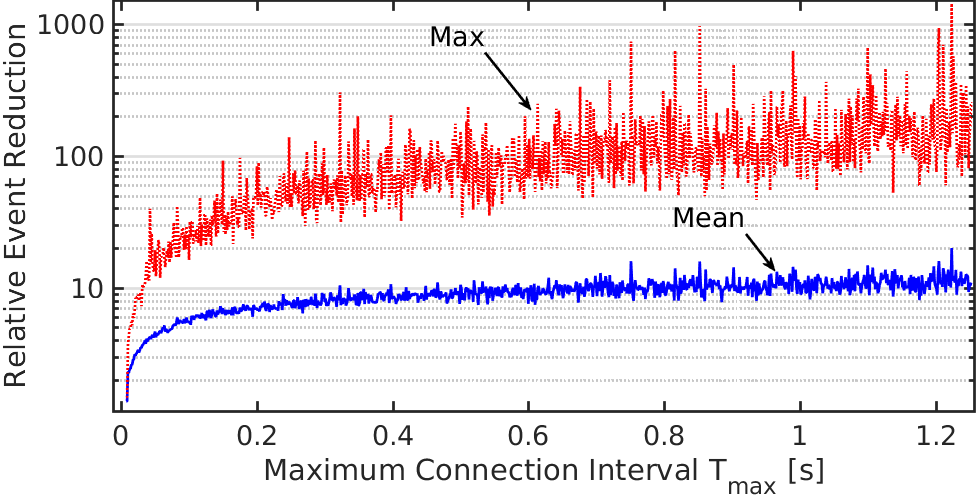}
\vspace*{-0.6cm}
\subcaption{Reduction of the number of simulated events.}
\label{fig:ex4b_event_reduction}
\end{subfigure}
\begin{subfigure}[b]{0.49\textwidth}
\includegraphics[width=1.0\linewidth]{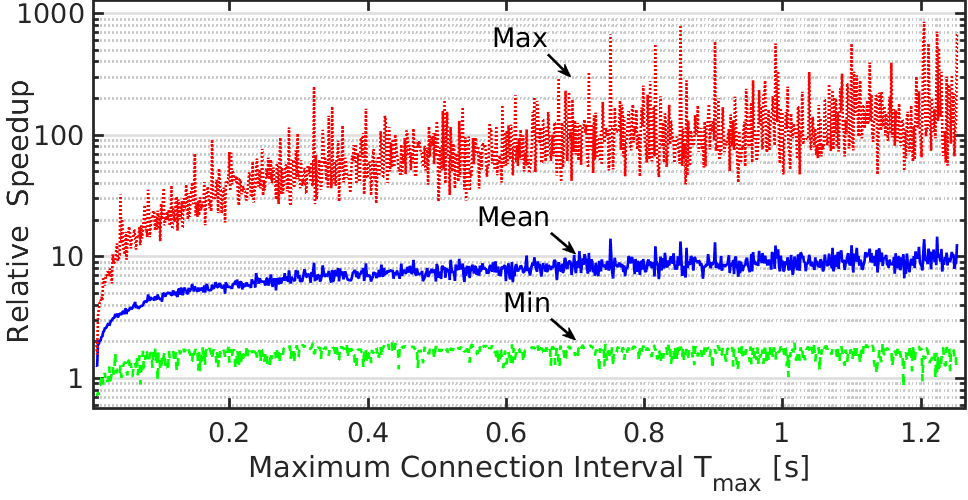}
\vspace*{-0.6cm}
\subcaption{Relative simulation speedup.}
\label{fig:ex4b_time_speedup}
\end{subfigure}
\end{center}

\vspace*{-0.2cm}

\caption{Event reduction and speedup for different values of $T_{max}$.\vspace*{-0.5cm}}
\label{fig:exp4b_speedup}
\end{figure*}

In this Section, we discuss the resulting collision rates, speedups and event reductions for 3 interfering networks. Each experiment has been repeated for $N_{r} = 200$ times. The number of used channels is 2, which is the minimal number of channels for BLE. The NoI is network $0$, which is the network with the largest connection interval.
\begin{figure}[tbh]
\begin{center}
\includegraphics[width=1.0\linewidth]{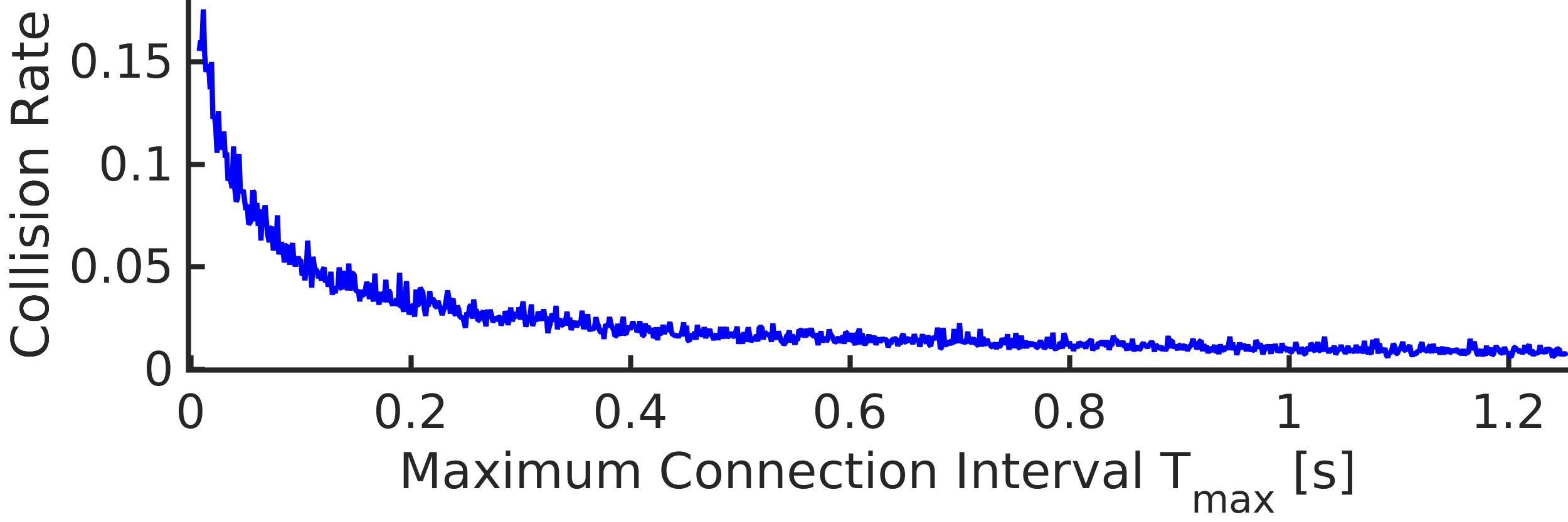}
\caption{Collision rate for different values of $T_{max}$.\vspace*{-0.3cm}}
\label{fig:ex4b_collision_rate}
\end{center}
\end{figure}
Figure \ref{fig:ex4b_event_reduction} depicts the mean collision rate simulated for this experiment. For each connection interval, the average value out of the 200 simulation repetitions has been computed. Due to the low number of channels used, for short connection intervals, the collision rate reaches almost $20 \%$.  For larger intervals, as expected, it gradually decreases.
The sum of simulation time in this experiment is $17.75$ hours without the speedup and $\SI{4.8}{h}$ with adaptive event skipping.
Figure \ref{fig:ex4b_event_reduction} depicts the relative event reduction, defined as the number of simulation events processed without adaptive event skipping divided by the number of events processed with adaptive event skipping. The figure shows the mean and maximum reductions of the number of events within the 200 repetitions for each connection interval. Higher reductions are achieved for larger values of $T_{max}$. Larger values of $T_{max}$ require more CPU computation as the simulation duration defined by Equation \eqref{eq:dsim} is increased such that more events need to be simulated. However, the larger the variance of simulated connection interval becomes, the higher the fraction of skipped events gets and hence the relative number of simulated events decreases thanks to the proposed adaptive event skipping. 

A similar result is shown in Figure \ref{fig:ex4b_time_speedup}, which depicts the time needed per simulation without adaptive event skipping divided by the time per simulation using our proposed speedup. For larger connection intervals, the mean approaches a value of 10, indicating that the simulation time is reduced by one order of magnitude. The maximum speedup is typically another order of magnitude above the mean speedup. For some especially beneficial subcases, it almost reaches a factor of 1000.
The minimum achievable speedup is below 1 for the case where $T_{max} \approx T_{min}$. This means that for very unfavorable subcases, the simulation is slowed down due to the overhead for the predictions. The minimum speedup that has been observed is $0.7083$. For all values of $T_{max} > 96.25~ms$, the minimum observed speedup is always above $1$. 
 
\subsubsection{Impact of the NoI}
\begin{figure}[tbh]
\begin{center}
\includegraphics[width=1.0\linewidth]{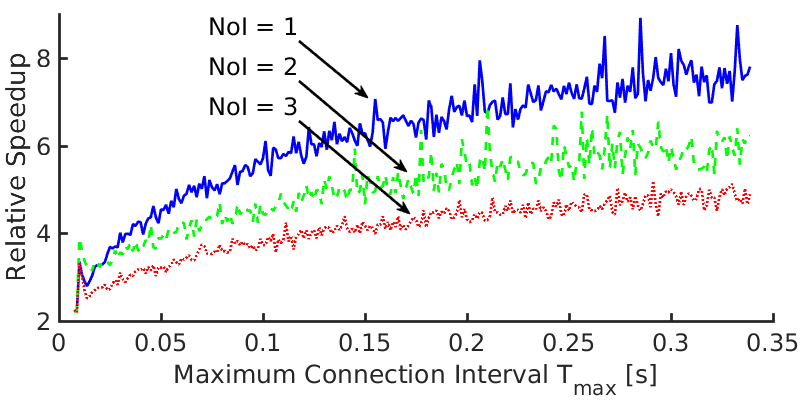}
\caption{Mean speedup when varying the index $n_{oI}$ of the network of interest.\vspace*{-0.5cm}}
\label{fig:ex1a-c_noi}
\end{center}
\end{figure}
The selection of the NoI also influences the achievable speedup. Recall that all connection intervals are chosen randomly. We regard $n_{oI}$ as the index of the NoI in a list of networks sorted in the descending order of their connection intervals. In this experiment, we investigate the effect of $n_{oI}$. Values larger than 1 effectively reduce the connection interval of the NoI relative to the intervals of the other networks. Since shorter connection intervals of the NoI lead to a higher number of collisions, less events can be skipped for all networks. We evaluate how the speedup changes according to this effect. In this experiment, 3 networks have been assumed, 1000 repetitions per connection interval value have been carried out and 37 channels have been used. Figure \ref{fig:ex1a-c_noi} shows the speedups for all three different values of $n_{oI}$. For $n_{oI} = 1$, the speedup has the highest value. For larger values of $n_{oI}$, it is reduced. The sums of simulation times (without/with adaptive event skipping) in this experiment are as follows: $\SI{30.8}{h}$/$\SI{8.0}{h}$ for $n_{oI} = 1$, $\SI{30.1}{h}$/$\SI{8.9}{h}$ for $n_{oI} = 2$ and $\SI{29.9}{h}$/$\SI{10.6}{h}$ for $n_{oI} = 3$. 

\subsubsection{Impact of the Number of Networks}
In this experiment, we evaluate the impact of different numbers of interfering networks on the achievable speedup. Network 1 is considered as the NoI. The number of channels has been set to 2, the simulation has been repeated 1000 times per connection interval.
\begin{figure}[bt]
\begin{center}
\includegraphics[width=1.0\linewidth]{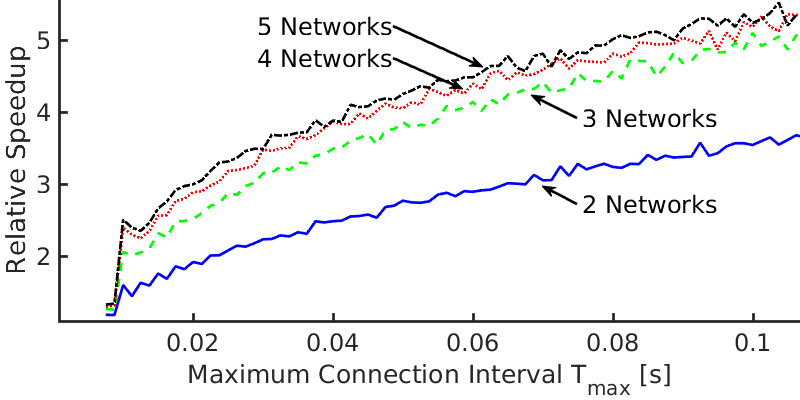}
\caption{Mean speedup for differnt numbers of colliding networks.\vspace*{-0.2cm}}
\label{fig:exp2a-d_nnetworks_1}
\end{center}
\end{figure}
As can be seen in Figure~\ref{fig:exp2a-d_nnetworks_1}, the number of networks influences the speedup achieved. The more networks are under simulation, the higher the mean speedup becomes. This property is especially beneficial because the speedup grows with increasing complexities of the problem studied. 
The total sum of simulated times (without/with our proposed speedup) in this experiment are: For 2 networks: $\SI{7.5}{s}$/$\SI{2.7}{s}$; For 3 Networks: $\SI{3.7}{min}$/$\SI{1.1}{min}$; For 4 networks: $\SI{1.5}{h}$/$\SI{0.4}{h}$; For 5 networks: $\SI{30.7}{h}$/$\SI{8.4}{h}$.
%

\subsection{Potential Side-Effects}
In the simulation model described in Section \ref{sec:simulationModel}, it is assumed that the slave always sends a packet irrespective of the preceding packet from the master has been received successfully or not. However, there is one case in which the response packet from the slave might not be sent.
The first byte of every packet consists of a preamble, which is used by the receiver for frequency synchronization and gain calibration \cite{bleSpec4.2}. If this preamble is lost at the packet from the master to the slave, the slave might miss the packet and not send any response. 
The slave's behavior in case of a preamble loss is not clearly defined by the BLE standard \cite{bleSpec4.2} and depends on the implementation.
\begin{figure}[htb]
\centering
\includegraphics[width=\linewidth]{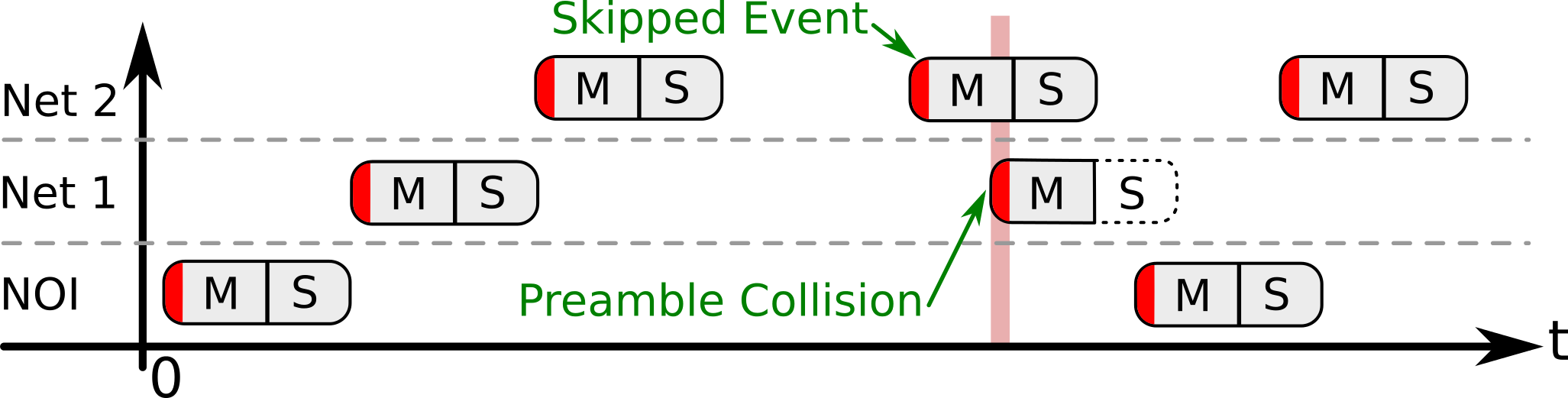}
\caption{Effect of colliding preambles.}
\label{fig:preambleCollision} 
\end{figure}

Figure \ref{fig:preambleCollision} depicts a situation with such side effects.
If a preamble is lost in the second connection event of Net1, the slave does not respond and hence it will not collide with the second connection event of NoI.
However, as second connection event of Net2 is skipped by adaptive event skipping, the simulator is not aware of the preamble loss and a collision is reported between Net1 and NoI.
Therefore, for protocols like BLE and stacks that are vulnerable to preamble loss, the collision rate obtained by our accelerated speedup is an upper bound of the actual one. 
For networks such as ANT/ANT+ \cite{AntSpec:14}, which support cyclic unidirectional packets from one node to another without requiring a response, this effect does not appear and the simulated collision rate is equivalent to the actual one. 
One way to overcome this effect are backtracking methods: For each colliding event, previously skipped ones are checked for overlapping preamble transmissions, until one with no overlap has been found. Based on these checks, the classification of the event as ``colliding'' can be revoked.

\section{Concluding Remarks}
\label{sec:concluding_remarks}
We have presented a speedup technique based on skipping events adaptively, which is capable of reducing the wall-clock time needed for conducting collision simulations of realistic situations by one order of magnitude. Our proposed technique allows the simulation of much more complex situations  within the same amount of time. For example, the number of repetitions in Monte Carlo simulations or the number of networks can be increased significantly compared to conventional discrete event simulations. We have shown and evaluated the proposed technique in the context of collisions among multiple BLE networks. However, adaptive event skipping is a generic technique and can also be used for other network protocols such as e.g. ANT/ANT+, as long as they are organized in a cyclic manner. Besides that, there are applications outside the networking domain. For example, it can be used for simulating the mean delay when multiple cyclic tasks need to be executed on the same processor, etc. In our future work, we plan to integrate backtracking mechanisms to account for potential side-effects in case of preamble collisions, as already described. In the context of Bluetooth Low Energy, we expect that our simulation technique will help to parametrize the protocol appropriately by accounting for collision probabilities. 


\bibliographystyle{IEEEtran}
{
	\tiny
	\bibliography{paper}
}
\end{document}